\documentclass[]{spie}  

\usepackage{amsmath, amsfonts, amssymb}
\usepackage{graphicx, siunitx, multirow, overpic}
\DeclareSIUnit \Msun {M_{\odot}}
\usepackage[colorlinks=true, allcolors=black]{hyperref}

\title{SuperFIRE: Concept evolution of a seeing-limited\\ broadband spectrograph for the GMT}

\author[*, a]{Gustav M. Pettersson}
\author[a, b]{Gábor Fűrész}
\author[a]{Nathan P. Lourie}
\author[a]{F. Elio Angile}
\author[a]{Jill Juneau}
\author[a]{Gerardo Berlanga Molina}
\author[a, c]{Robert A. Simcoe}
\affil[a]{MIT Kavli Institute for Astrophysics and Space Research, Cambridge, MA, USA}
\affil[b]{Schmidt Sciences - Astrophysics \& Space Center, New York, NY, USA}
\affil[c]{Department of Physics, Massachusetts Institute of Technology, Cambridge, MA, USA}

\authorinfo{*Corresponding author: gupet@mit.edu}

\begin{document} 
\maketitle

\begin{abstract}
We are developing the SuperFIRE concept, which builds on the heritage of the FIRE spectrograph at Magellan, as a broadband (about \SI{340}{\nano\m} to \SI{2.5}{\micro\m}) intermediate resolution (R=10,000) single-shot natural seeing spectrograph for the Giant Magellan Telescope (GMT) with support from the Kavli Foundation. This single-object spectrograph is envisioned as a first-light instrument that can operate in natural seeing and grow in long-term capability as the telescope and adaptive optics systems mature. Here we discuss how the science cases and design have evolved since the concept was initially conceived and presented at SPIE in Edinburgh ten years ago. The primary drivers for the design changes are fast follow-up of faint transients and multi-messenger events, and complementarity to JWST observations in the infrared. By leveraging fast low-noise detectors and modern fabrication techniques we expect high throughput and low scatter that delivers sky limited performance in a few minutes of integration time for fast response and deep integrations. SuperFIRE will operate as a practical point-and-shoot follow-up spectrograph thanks to a versatile standard configuration that needs minimal observing or planning overheads and is insensitive to atmospheric conditions. 
\end{abstract}


\section{INTRODUCTION}
\label{sec:intro}  
The concept of a ``SuperFIRE'' spectrograph for the for the \SI{25.4}{\m} Giant Magellan Telescope (GMT) based on the popular Folded-port InfraRed Echellette (FIRE) \cite{simcoeFIREFacilityClass2013} spectrograph at the \SI{6.5}{\m} Magellan Baade telescope was first proposed ten years ago at SPIE in Edinburgh.\cite{simcoeConceptSeeinglimitedNearIR2016} The original motivation for SuperFIRE was to enable impactful near-infrared (NIR) science early in the phased construction of the GMT with an affordable instrument operating in natural seeing, as a compliment to subsequent adaptive optics (AO) assisted NIR instrumentation (i.e.\ GMTNIRS\cite{jaffeGMTNIRSProgressGiant2016a} and GMTIFS\cite{sharpGMTIFSGiantMagellan2016}). That version of SuperFIRE emphasized low risk through mature off-the-shelf technologies and didn't plan for capability evolutions such as the ground-layer adaptive optics (GLAO)\cite{bouchezGiantMagellanTelescope2014} system that the telescope will deliver. The original concept study pointed out that in the future SuperFIRE could be a powerful tool for emerging transient science cases and that it would be especially powerful if it could be operated in parallel with the optical spectrograph GMACS\cite{fabricantGMACSModeratedispersionOptical2025}. Today we see both transient science and broadband UV-optical-NIR coverage as essential parts of the SuperFIRE science case and have evolved the concept accordingly. Note that SuperFIRE remains a concept that has not yet been adopted as an instrument in the GMT suite.

Through a concept study supported by the Kavli Foundation we have identified a need for fast broadband (UV through NIR) follow-up capability of transients discovered by current and near-future deep-wide-fast surveys (e.g.\ Rubin and Roman) and multi-messenger opportunities from upgraded and future gravitational wave observatories (e.g.\ LIGO and LISA) that are too faint to characterize with current generation spectrographs. While natural-seeing or GLAO instruments have lower ultimate sensitivity due to increased sky background they can be designed to get on-target and integrating quickly and with minimal planning which gives a fast response and efficient use of valuable telescope time. To meet this emerging science need the new SuperFIRE concept has been extended to include include UV-optical spectral coverage alongside NIR, has enhanced sensitivity and flexibility through novel technologies, and has been optimized for shorter integration times than the original concept. 

The new concept for SuperFIRE is a modular instrument with two white-pupil echelle spectrographs, one at ambient temperature for the UV-optical (\SIrange{.34}{.95}{\micro\m}) and one at cryogenic temperature for the NIR (\SIrange{.95}{2.5}{\micro\m}), with a shared front-end that provides alignment and guiding. The NIR portion is a direct evolution of the original concept with increased spectral resolution to $R=10{,}000$ from $R=6000$ and a reduced beam size and detector area. These improvements are enabled by replacing the nominal .7" slit with a 3x.24" image slicer to reduce the slit-resolution product of the spectrograph and allow better coupling to the range of expected image quality delivered across atmospheric conditions, with and without GLAO, so that a fixed configuration can be used for most observations. The baseline design for the UV-optical segment is carried over from the NIR concept to reduce development complexity and time, but we are actively exploring other possibilities during the current concept study. SuperFIRE is designed to be a simple and efficient point-and-shoot spectrograph that captures its broadband spectral range in a single shot following the model pioneered by X-Shooter\cite{vernetXshooterNewWide2011}. In addition, we require the instrument to always be ready to start science integrations within five minutes and will provide quick-look reduced data within two minutes of an exposure's end to maximize responsiveness for dynamic observing of transient science cases, and specify the use of frame-transfer detectors to practically eliminate any dead-time between exposures.

The new SuperFIRE design remains small and affordable for an extremely large telescope instrument and is designed for the instrument platform (IP) station on the GMT. At its intermediate resolution SuperFIRE will be a complement to the strong suite of high-resolution single-object spectrographs already planned for the GMT, i.e.\ G-CLEF that covers \SIrange{.35}{.95}{\micro\m} at $R=19{,}000 \text{--} 105{,}000$\cite{szentgyorgyiGMTconsortiumLargeEarth2018} and GMTNIRS that covers \SIrange{1.1}{5.4}{\micro\m} at $R\geq65{,}000$\cite{leeGMTNIRSOpticalSystem2022}, each in a single shot. It also complements GMTIFS\cite{sharpGMTIFSGiantMagellan2016} and GMACS\cite{fabricantGMACSModeratedispersionOptical2025} by providing similar resolution and sensitivity but giving observers a choice between a high spatial resolution NIR integral field unit/imager (GMTIFS), wide-field multiplexing in UV-optical (GMACS), or single-object broadband coverage (SuperFIRE).

\section{DRIVING SCIENCE CASES}
\subsection{Faint-object spectroscopy}
SuperFIRE's capabilities of NIR faint-object spectroscopy that were highlighted in the original paper\cite{simcoeConceptSeeinglimitedNearIR2016} remain important science drivers for the instrument design but recent work by the James Webb (JWST) and Euclid space telescopes, and expectations from the upcoming Nancy Grace Roman Space Telescope, motivate upgrades to the SuperFIRE instrument concept. The so-called little red dots (LRDs) that were discovered in early JWST observations may be fundamental to understanding supermassive black hole and galaxy formation in the early universe and exhibit ``V-shaped'' spectral energy distributions with significant rest-frame UV and NIR emission\cite{settonLittleRedDots2025} that require deep and broadband optical through IR spectroscopy to study. JWST provides excellent $R\approx2700$ spectroscopy past about \SI{1}{\micro\m}, which SuperFIRE will complement by reaching similar limiting magnitudes with significantly higher resolution, and filling in the data below \SI{1}{\micro\m}. Euclid and Roman will also provide a large sample of distant quasars to study the epoch of reionization with SuperFIRE, with samples of quasars at $J\gtrsim23$ already known from Euclid\cite{yangEuclidDiscovery312026} and completeness to $J\approx25.5$ expected with Roman\cite{teePredictingYields652023}. With UV-optical coverage SuperFIRE will also be used to study faint stars in the outskirts of the Milky Way dark-matter halo down to the main sequence turnoff, to enhance our understanding of the local universe's assembly and dark matter structures. Delivering the performance necessary to capitalize on these science cases motivates the increased resolution and bandwidth, strict stray light and sky subtraction, and GLAO compatibility requirements of the updated SuperFIRE concept.

\subsection{Time domain and multi-messenger transients}
The importance of time-domain and multi-messenger astronomy was reaffirmed in the Astro2020 decadal survey \cite{committeeforadecadalsurveyonastronomyandastrophysics2020astro2020Astro2020PathwaysDiscovery2023} and builds on major public and private investment in this field in the last few decades. With the arrival of the Rubin Observatory on the ground and Roman in space we are entering a follow-up limited era with thousands of transients across optical and NIR discovered each day without the capability to spectroscopically study a significant fraction of them. The scientific appetite for follow-up capability is clear from from these proceedings, with projects including SOXS\cite{SOXS}, NTE\cite{NTE}, NINJA\cite{NINJA}, and ZShooter\cite{ZShooter} reaching commissioning or rapidly advancing their development. On a dedicated \SI{3}{m}-class telescope, SOXS will routinely provide $R\approx5000$ follow-up down to magnitude $r\approx20$\cite{schipaniSOXSWideBand2018}, but the bulk of discovered transients will be several magnitudes fainter and require the collecting area of an extremely large telescope for spectroscopic follow-up.

Studying transients and leveraging multi-messenger data is fundamental to understand several open questions in astrophysics. For example, the timeline and mechanisms of heavy element formation have long been postulated but only recently been possible to study thanks to the gravitational wave discovery of the binary neutron star merger GW170817 and its kilonova follow-up across the electromagnetic spectrum \cite{cowanOriginHeaviestElements2021}. The merger GW170817 peaked at an absolute magnitude of around -16 across optical-NIR, with its peak flux moving from blue within hours of the merger to K-band after about a week,\cite{arcaviFirstHoursGW1708172018} which highlights the broadband nature of these events. At a distance of about \SI{40}{Mpc} its apparent magnitude peaked around 17 which was possible to localize and follow up with contemporary observatories. Since GW170817 the gravitational waves of only one other potential binary neutron star merger have been detected \cite{abbottGW190425ObservationCompact2020} but its greater distance (c.\ \SI{159}{Mpc}) and large position uncertainty made electromagnetic follow-up unsuccessful. To continue the work started with GW170817 we need both more sensitive gravitational wave observatories and deeper electromagnetic follow-up capabilities.

In observing run 5, LIGO is expected to reach a binary neutron star merger range over \SI{300}{Mpc} and uncover tens of kilonovae whose apparent magnitude on discovery would be $r\gtrsim23$\cite{shahPredictionsElectromagneticCounterparts2024}, which is well matched to fast localization with Rubin but will be difficult to follow-up spectroscopically with current facilities. Beyond multi-messenger events, Rubin and Roman will discover an unprecedented quantity and diversity of transients, the majority of which will be near their sensitivity limits as volume scales with distance cubed. Spectroscopic follow-up of optical magnitudes $r\gtrsim24$ and infrared magnitudes $J\gtrsim25$ are therefore necessary to be able to classify and understand these transients, which in practice is beyond the reach of current spectroscopic facilities except long integrations with JWST. To fill this capability gap a driving science requirement for SuperFIRE is to be able to spectroscopically classify any Rubin $r=24$ transient in less than one hour, and any southern Roman $J=25$ transient in less than five hours.

\section{INSTRUMENT REQUIREMENTS}
\subsection{Wavelength coverage, resolution, and slit selection}
\begin{table}[htb]
    \centering
    \caption{Performance specifications considered in the SuperFIRE concept design study.}
    \vspace{2mm}
    \begin{tabular}{c|ccc|c}
        Parameter & Requirement & Goal & This concept & Old concept \\\hline
        Resolution & $R\geq6000$ & $R=30{,}000$ & $R=10{,}000$ & $R=6000$ \\
        Bandwidth & \SIrange[range-phrase = --, range-units = single]{.360}{2.40}{\micro\m}
                  & \SIrange[range-phrase = --, range-units = single]{.325}{2.51}{\micro\m}
                  & \SIrange[range-phrase = --, range-units = single]{.340}{2.47}{\micro\m}
                  & \SIrange[range-phrase = --, range-units = single]{.87}{2.5}{\micro\m}\\
        Slit width & .72"-.32" & .86"-.25" & 3x.24" = .72" & .7" nom.\ \\
        Slit length & $6"$ & $10"$ & 15" virtual & 8"\\\hline
        \multirow{2}{*}{Low-R mode} & \multirow{2}{*}{none} & 1' longslit & 15" slit & 15" slit \\
         &  & at $R\approx1000$ & at $R=3300$ & at $R=800$
    \end{tabular}
    \label{tab:reqs}
\end{table}
The requirements for SuperFIRE have evolved from FIRE and the original SuperFIRE concept in three main areas: i) adding coverage of the UV-optical as a broadband single-shot spectrograph, ii) a strong desire to increase the spectral resolution, and iii) optimizing the design for long-term GLAO performance as a complement to natural seeing operation. A summary of the specifications that we explored in the current concept study are given in Table~\ref{tab:reqs}. Because it is difficult to realize one spectrograph that covers the whole wavelength range effectively SuperFIRE will be split into two spectrographs, and therefore only a portion of these requirements will flow down to each subsystem. We have limited the design to two spectrographs with multiple camera arms after also considering a three spectrograph solution like X-Shooter\cite{vernetXshooterNewWide2011} and finding the volume of such a solution to scale poorly to the physical size of the GMT.

The resolution and slit dimension requirements stem from the original design of FIRE, where keeping $R\geq6000$ and a slit length of at least 6" are fundamental to resolve the atmospheric OH-lines and perform good sky subtraction to observe faint targets in the NIR. To deliver consistent data for broadband science cases we require the same resolution across the UV-optical as well. We set strong goals to increase the slit length and spectral resolution to at least $R=10{,}000$ for better sky subtraction and reduced influence from atmospheric emission lines, as the targets for SuperFIRE will be faint compared to the background. Because of the atmosphere's line emission in the NIR we expect to achieve better sensitivity by recording a higher resolution spectrum and binning it (say to $R\approx1000$) after sky subtraction, than by capturing at a native lower resolution. We found that a resolution of $R=30{,}000$ would be ideal to suppress atmospheric line emission and enhance our stellar spectroscopy science cases, but that $R=20{,}000$ was the maximum feasible with two spectrographs and chose $R=10{,}000$ for the baseline as it halved the estimated cost of the instrument without any dramatic loss in the GMT's science potential thanks to its existing strong suite of high-resolution spectrographs.  The slit width requirement was broadened from FIRE's .6" nominal value to a range of .72" to .32" to include UV-optical and be compatible with both NS and GLAO operational modes.

The bandwidth limits are derived from science cases that need to measure Ly-$\alpha$ at a Cosmic Noon redshift of $z=2$ (c.\ \SI{365}{\nano\m}) and the CO band-head overtone up to \SI{2.4}{\micro\m} respectively. Coverage from \SIrange{.325}{2.51}{\micro\m} was set as the goal as that would encompass the entire atmospheric window up to an extinction of 1 magnitude. We believe that reaching far into the UV would expand SuperFIRE's scientific capabilities but also drives up the cost and difficulty in realizing the instrument. Ultimately, the UV throughput of the telescope places practical limits on the potential performance and we chose the baseline value of \SI{340}{\nano\m} to allow several i-line glasses in the optical design and match the cutoff of a Gemini-style UV enhanced silver coating.

As an optional feature, we would like to replicate FIRE's popular low-resolution high-throughput mode that provides $R\approx500$ with a 50" longslit in SuperFIRE. In FIRE the main disperser is mounted on a turret that allows it to be swapped for a mirror so only the cross disperser is used to generate spectra. The original SuperFIRE concept included a similar turret feature to give an $R\approx800$ mode, but a slit longer than about 15" could not be accommodated. Given that the new SuperFIRE concept has a native 15" slit we can remove the complexity of a grating turret and instead use an unsliced physical slit to get a low-resolution mode at $R\approx3300$ for science cases that require spatial information. The full resolution will also be available with a physical .24" by 15" slit.

\subsection{Detector performance and stray light}\label{sec:stray_light}
As a natural seeing/GLAO instrument on an extremely large telescope, SuperFIRE's performance will strongly depend on the intensity of background sources and, in the NIR especially, our ability to suppress stray light originating from bright atmospheric lines. In the new SuperFIRE concept we have increased our sensitivity and speed goals significantly, which requires fast low-noise detectors and about an order of magnitude lower stray light in the instrument. We define detector performance requirements so that their noise contribution is small relative to the lowest realistic background and introduce strict stray light requirements in the NIR to ensure that bright lines are controlled. Under these conditions SuperFIRE will perform near the physical limits of what is possible for any spectrograph that operates through Earth's atmosphere.

\begin{figure}[p]
    \centering
    \includegraphics[width=0.7\linewidth]{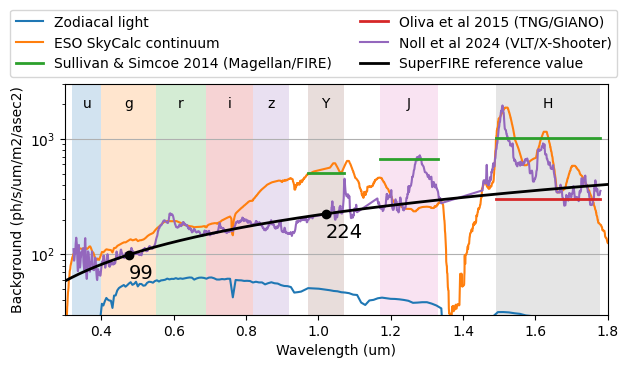}
    \caption{Continuum background spectral radiance used to derive SuperFIRE requirements. The value for zodiacal light is taken at 1.2 times the minimum. The SkyCalc (\url{https://www.eso.org/observing/etc/skycalc/}) and Noll et al.\cite{nollStructureVariabilityOrigin2024} values were scaled by 1.2 to represent typical observing conditions and have zodiacal light added. The Sullivan \& Simcoe\cite{sullivanCalibratedMeasurementNearIR2012} and Oliva et al.\cite{olivaLinesContinuumSky2015} values are given as typical values for the respective bands.}
    \label{fig:background}
\end{figure}
\begin{figure}[p]
    \centering
    \includegraphics[width=0.7\linewidth]{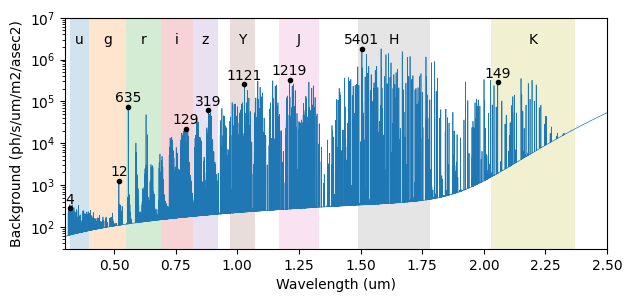}
    \caption{Total background model for deriving requirements, consisting of the reference continuum, lines sourced from SkyCalc, and thermal emission. The brightest line in each band is marked with its value relative to the continuum.}
    \label{fig:bg_with_lines}
\end{figure}

It is challenging to define the continuum sky background as it has a complex structure that varies over time\cite{nollStructureVariabilityOrigin2024} and has proven difficult to measure accurately with current instrumentation, especially in the NIR\cite{sullivanCalibratedMeasurementNearIR2012, viuhoNearinfraredAirglowContinuum2025}. Zodiacal light, scattered moonlight, and thermal emission from the telescope mirrors are also significant but well understood contributors to the background. We estimate a simple but realistic continuum background for SuperFIRE as linearly increasing from \SI{99}{ph/s/\micro m/m^2/asec^2} in g-band to \SI{224}{ph/s/\micro m/m^2/asec^2} in Y-band, based on Noll et al.'s VLT/X-Shooter data\cite{nollStructureVariabilityOrigin2024} plus zodiacal light. This continuum background is shown in Figure~\ref{fig:background} compared with available measurements to illustrate the uncertainty. The complete background model includes atmospheric emission lines and thermal emission for a telescope with an emissivity of 10\% at \SI{10}{\celsius} and is shown in Figure~\ref{fig:bg_with_lines}.
\begin{table}[p]
    \caption{Derived nominal detector requirements to reach background dominated sensitivity. The effective QE is the actual QE divided by the excess noise factor for amplified detectors (e.g.\ EMCCD and LmAPD).}
    \centering
    \begin{tabular}{r|ccccc|cccc|c}
        Band & \bf u & g & r & i & z & \bf Y & J & H & K \\\hline
        RN (eRMS) & \bf .72 & .97 & 1.28 & 1.56 & 1.81 & \bf 1.10 & 1.35 & 1.79 & 8.4 & \multirow{3}{*}{Required} \\
        DC (e/px/h) & \bf 1.0 & 1.9 & 3.3 & 4.9 & 6.5 & \bf 3.6 & 5.0 & 10 & 211 \\
        Effective QE & \bf 83\% & 83\% & 83\% & 83\% & 83\% & \bf 69\% & 69\% & 69\% & 69\%\\\hline
        RN (eRMS) & \bf .59 & .79 & 1.04 & 1.28 & 1.48 & \bf .67 & .83 & 1.09 & 5.1 & \multirow{2}{*}{Goal} \\
        Effective QE & \bf 91\% & 91\% & 91\% & 91\% & 91\% & \bf 76\% & 76\% & 76\% & 76\%\\
    \end{tabular}
    \label{tab:detector_req}
\end{table}

Our overall sensitivity requirement is to produce spectra with a signal-to-noise ratio (SNR) that is at least $\frac{2}{3}$ (with a goal of $\frac{3}{4}$) of the background-only SNR for a \SI{5}{min} reference exposure. This short exposure time is chosen to suit the transient science cases, improve sky subtraction, and avoid significant saturation from sky lines. With our sensitivity requirement the shot noise due to stray light and signal degradation due to detector quantum efficiency (QE), read noise (RN), and dark current (DC) can raise the noise floor by no more than 50\% over the reference background, of which we allocate 30\% to the detectors, 10\% to stray light, and hold 10\% as margin. With some notional assumptions on the spectrograph's parameters such as throughput (30\%) and pixel size (.1"/px) we derive the nominal detector requirements shown in Table~\ref{tab:detector_req}. Notably, even with an order of magnitude increase in collecting area over current telescopes, SuperFIRE would be read noise limited with current generation detectors, e.g.\ with a typical 3 eRMS CCD in UV-optical and a 4 eRMS HxRG in NIR. Novel detector technologies are therefore key to enable SuperFIRE's fast follow-up capability and potentially also faint science cases if the sky subtraction improves with shorter exposures as we expect. We are actively engaged in a series of trade studies and laboratory test campaigns to evaluate promising detector technologies. These investigations are currently focused on skipper and/or SiSeRO CCDs\cite{fernandezmoroniSubelectronReadoutNoise2012,chattopadhyayDemonstratingSubelectronNoise2024} or next generation qCMOS\cite{Lightspeed} detectors for the UV-optical and linear-mode avalanche photodiode (LmAPD)\cite{claveauProgressMegapixelLinearmode2024} arrays for the NIR spectrograph, but our timeline and design allows us to be responsive to new detector technologies that emerge as the project matures.

To meet the stray-light noise contribution requirement SuperFIRE will require an extremely dark (about a factor of 10 darker than FIRE) spectrograph and extremely low scattered light. Attention to detail and significant baffling will be necessary in the former case to ensure there are no emitters in the spectrographs and that the detectors are protected from external radiation. FIRE is known to be scattered-light limited in H-band\cite{sullivanCalibratedMeasurementNearIR2012} and sees a background that is about three times higher than the continuum (see Figure~\ref{fig:background}). We believe the main sources of scatter are the echelle grating surface roughness and the internal grain boundaries of the zinc selenide cross dispersion prisms (see Ikeda et al.\cite{ikedaZincSulfideZinc2009}), and therefore that novel grating manufacturing techniques are a key technology for SuperFIRE and that careful glass selections will be important. We have allocated half of the stray light budget to diffuse scattered light from atmospheric lines and determined that the H-band total scatter needs to be suppressed to a level of $2\times 10^{-3}$ (with a goal to suppress the brightest line to the continuum level, i.e.\ by $2\times 10^{-4}$). As an example, we have allocated $\frac{1}{5}$ of the total scatter to the echelle grating, and therefore require a surface roughness below \SI{2.6}{\nano\meter} (goal \SI{.8}{\nano\meter}) RMS.

\subsection{Operational efficiency}
Operational efficiency is an important metric for all instruments, and especially on the GMT scale where the cost per hour to build and operate the observatory is substantial. With SuperFIRE envisioned as a dynamically available instrument the efficiency becomes critical to minimize disruption to the observing queue when targets of opportunity appear. Our core philosophy when designing SuperFIRE for operational efficiency has been:
\begin{enumerate}
    \setlength{\itemsep}{0pt} 
    \item Minimize the need for configuration, calibration, and planning before an observation.
    \item Get on target as quickly as the telescope allows, and don't waste photons on readout or sky subtraction.
    \item Keep to simple high-throughput designs where feasible and push noise sources below the sky background.
    \item Ensure the quick-look pipeline provides observers with the right data to know when to end integrating.
\end{enumerate}

As discussed in Section~\ref{sec:optical_decisions} we have chosen to equip SuperFIRE with a 3x image slicer instead of a traditional slit. We believe this has an operational efficiency advantage as a slicer gives low slit losses and high resolution across a wide range of delivered image quality in a versatile ``default'' configuration. We have chosen a dual-slicer layout and maximized the separation between the slicers to use ABBA nodding where we always observe both the target and the sky background in a manner that minimizes systematic errors. SuperFIRE will also have a 2' field-of-view (FoV) guider to provide over $99\%$ sky coverage for offset tracking of faint targets near the south galactic pole and a capability of deep imaging and photometry to quickly localize targets with poor astrometry. Finally, we specify frame-transfer CCD or CMOS sensors so that the overhead between exposures is only a few seconds, which is especially important with the short exposure times we expect with SuperFIRE.

As a dynamic observing tool we also consider that the information and tools that SuperFIRE presents ``live'' to the observers are critical to help them decide when they have reached their desired data quality or how much they should prioritize the current target versus moving on or ``releasing'' the telescope back to the queue. We have set a goal that quick-look data should be presented no more than two minutes after an exposure ends, envisioning a default observing strategy of starting a $2\times5$ min AB sequence, and giving the observer the first data at 7 minutes with an option to end having only used 10 minutes of telescope time, or continuing an ABBA sequence until the desired data quality is reached.

\section{OPTICAL DESIGN}
\subsection{Challenges}
As a seeing-limited (or GLAO seeing-enhanced) spectrograph on an extremely large telescope, SuperFIRE faces three main challenges compared to fully AO-assisted instruments. First, in the seeing limit the resolution of a spectrograph is inversely proportional to the telescope's aperture, and so the beam size and therefore physical size of the instrument generally becomes large. GMT/SuperFIRE's aperture-resolution product is approximately seven times larger than that of Magellan/FIRE, but a seven times larger instrument is not practical so a straightforward scaling of FIRE is not feasible. Second, since the atmospheric seeing spreads the target's light over a larger area than diffraction alone, more of the sky's background scatter and emission must be let into the spectrograph entrance slit in natural seeing than with AO, which adds background noise. This background limits the ultimate sensitivity of the instrument and introduces stray light challenges but also means that if the instrument noise contribution is negligible compared to the background the resolution can be increased without a sensitivity penalty. Finally, it is optically challenging to implement compact natural seeing instruments at this scale as the final focal ratio needs to approach unity to get reasonable spatial sampling (f/1.2 gives around .1" per \SI{15}{\micro\m} pixel with the GMT).

\subsection{High-level decisions}\label{sec:optical_decisions}
SuperFIRE is based on a pair of white-pupil echelle spectrographs that respectively cover the UV-optical in an ambient design with silicon detectors and the NIR in a cryogenic design with mercury-cadmium-telluride detectors. To meet the strict background requirements any part of the instrument with detectors sensitive beyond about \SI{1.1}{\micro m} must be cryogenic and the dichroic split was placed between z and Y bands at \SI{.95}{\micro m} because this is where the highest QE transitions from Si to HgCdTe. Because the detector performance is critical and the final focal ratio must be fast we limit the design space to 2k \SI{15}{\micro m} detectors (or equivalent area) as larger detectors carry a development risk and make the cameras difficult to realize. In this context white-pupil echelle spectrographs are a natural choice to keep the camera aperture feasible and to match the spectral format with square detectors. In the UV-optical it may be possible to use very large detectors and binning to reach the desired spatial sampling and potentially a dichroic-tree first-order layout, see e.g.\ SOXS\cite{sanchezOpticalDesignSOXS2018}, as well.

As a typical rule-of-thumb, the slit width for a point-source spectrograph should be 1--1.5 times the full width at half maximum (FWHM) of the delivered image quality. A wider slit allows more light and a narrower slit gives higher resolution, so there is an inherent tradeoff and we consider around 1.2 FWHM to be an ideal slit width. The delivered image quality to SuperFIRE in median conditions is expected to range from about .72" to .48" in u to K band in natural seeing, and about .58"--.21" with GLAO. By using different slit widths for UV-optical and NIR it is possible to get good coupling across SuperFIRE's broad range in natural seeing, but with GLAO the image quality variation with wavelength is so steep that two slits are insufficient to guarantee good coupling and the performance would be sensitive to atmospheric conditions. Additionally, the natural seeing slits would project across 4--8 pixels so the detector area is not well used and the slit-resolution product of the spectrograph has to be high.

\begin{figure}[tbp]
    \centering
    \includegraphics[width=0.3\linewidth]{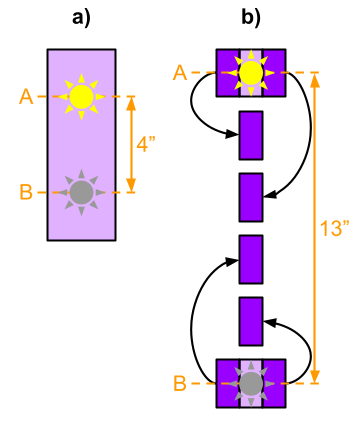}
    \caption{a) The .7"x8" physical slit of the old SuperFIRE concept. b) The .24"x15" virtual slit of the new SuperFIRE concept, which consists of two 3x.24" slicers with 2" long slitlets. Widths are exaggerated by a factor of four.}
    \label{fig:slicer_concepts}
\end{figure}
To overcome these limitations, we plan to use a 3x image slicer as shown in Figure~\ref{fig:slicer_concepts} to better couple to the range of image quality that SuperFIRE will see and to better match the projected slit width to the pixel scale. In particular, the chosen 3x.24" image slicer is simultaneously wide enough in total (.72") and narrow enough per slitlet (.24") that it will couple well to all conditions and wavelengths so that most observations can use the same configuration and produce consistent data. Additionally, we can now reach the ideal 2--3 pixels per projected slit width at desirable spatial sampling scales. It should be noted that in the case of SuperFIRE where a traditional slit would be oversampled there is no read-noise penalty introduced by the slicer, unlike situations where the pixel scale is made finer to sample a sliced slit.

As we discussed in Section~\ref{sec:stray_light}, novel echelle gratings are a key technology to mitigate scattered light in SuperFIRE and their inclusion has an additional benefit because we can freely specify their geometry, instead of choosing from a limited catalog of traditionally replicated options. Since the spectral resolution scales with the product of the beam diameter and the tangent of the blaze angle it is possible to build very compact high-resolution echelle spectrographs, but higher blaze angles increase the anamorphic magnification, i.e.\ projected slit width variation across each order, and therefore give an inconsistent resolution and spectral sampling. By defining a desired sampling of 2.1--3 pixels per projected slit, we find that a \SI{102}{mm} beam diameter, a \SI{55.5}{\degree} (R-1.46) blaze angle, and .098"/px is the ideal choice for an $R=10{,}000$ spectrograph with a .24" slit.

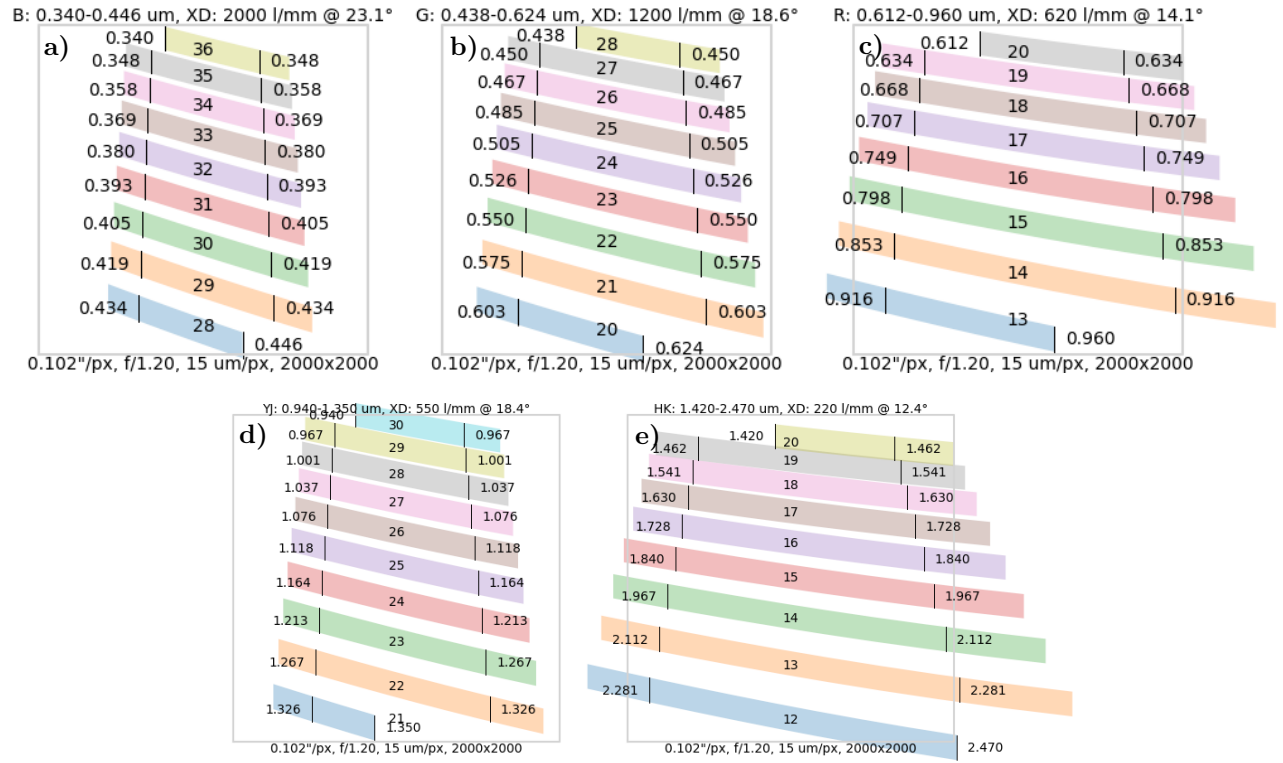
\begin{figure}[p]
    \centering
    \begin{overpic}[trim={15, 10, 630, 190}, clip, height=0.29\linewidth]{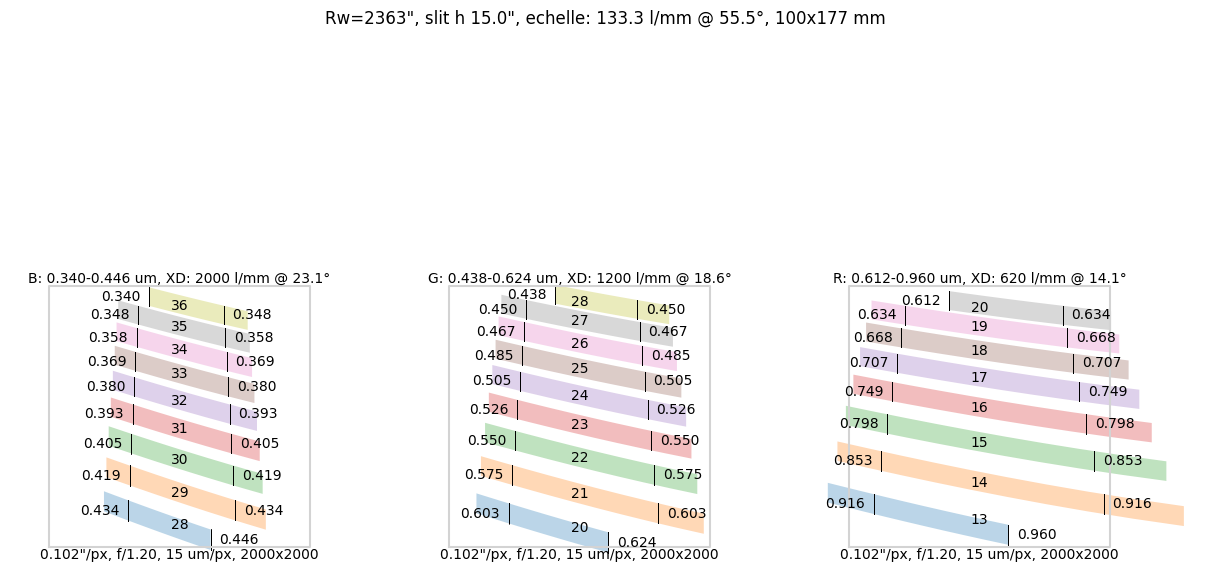}
        \put(10,80){\bf a)}
    \end{overpic}
    \begin{overpic}[trim={305, 10, 340, 190}, clip, height=0.29\linewidth]{UVIS_baseline.png}
        \put(10,80){\bf b)}
    \end{overpic}
    \begin{overpic}[trim={590, 10, 10, 190}, clip, height=0.29\linewidth]{UVIS_baseline.png}
        \put(9,67){\bf c)}
    \end{overpic}\\[3mm]
    \begin{overpic}[trim={48, 10, 548, 155}, clip, height=0.29\linewidth]{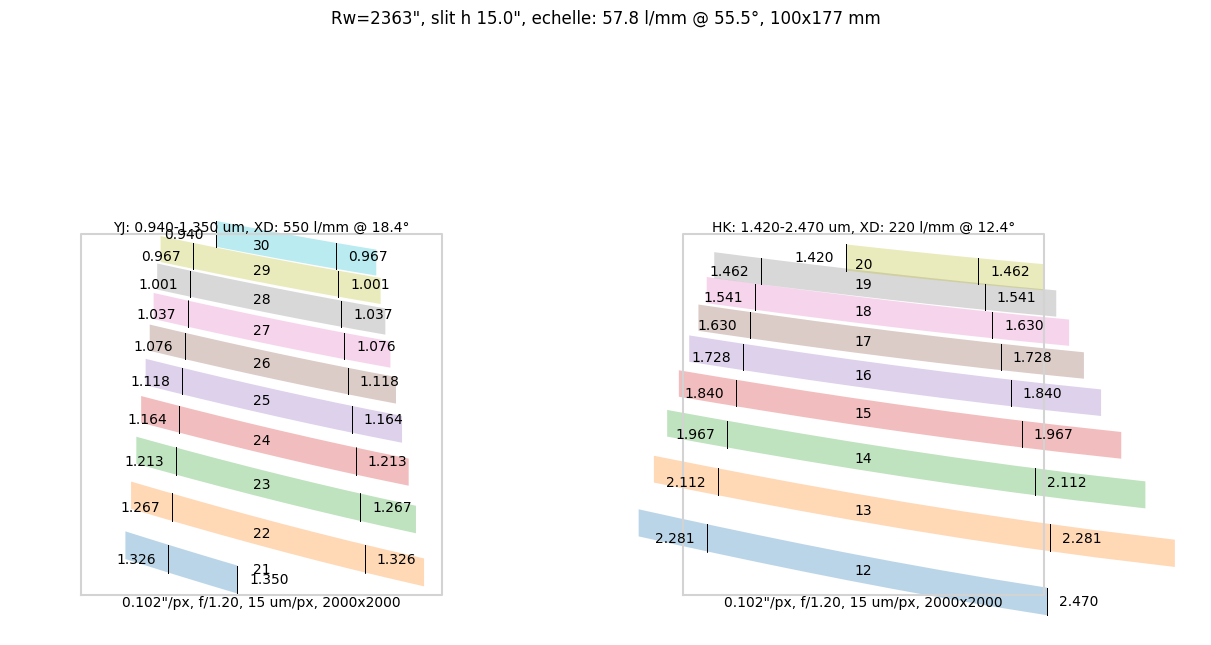}
        \put(5,88){\bf d)}
    \end{overpic}
    \begin{overpic}[trim={450, 10, 23, 155}, clip, height=0.29\linewidth]{NIR_baseline.png}
        \put(12,66){\bf e)}
    \end{overpic}
    \caption{SuperFIRE baseline echellograms for UV-optical (a--c) and NIR (d--e) with 1.5 FSR drawn in each order and the FSR bounds marked with vertical black lines and wavelengths in \si{\micro m}. These echellograms assume a \SI{100}{mm} beam and \SI{55.5}{\degree} blaze with a rule density of \SI{133.3}{lines/mm} and \SI{57.8}{lines/mm} respectively. The slit length is 15" and the minimum order spacing is about 10 (binned) pixels, except for d) which will receive additional cross dispersion from a prism.}
    \label{fig:parametric}
\end{figure}

The baselined parametric echellograms for the $R=10{,}000$ SuperFIRE concept are shown in Figure~\ref{fig:parametric} where we have rounded the ideal values and chose rule densities that place the \SI{.95}{\micro m} crossover at a blaze peak with a target of 1 and 1.2 free spectral range (FSR) coverage in K and z bands respectively.

\subsection{Spectrograph layout}
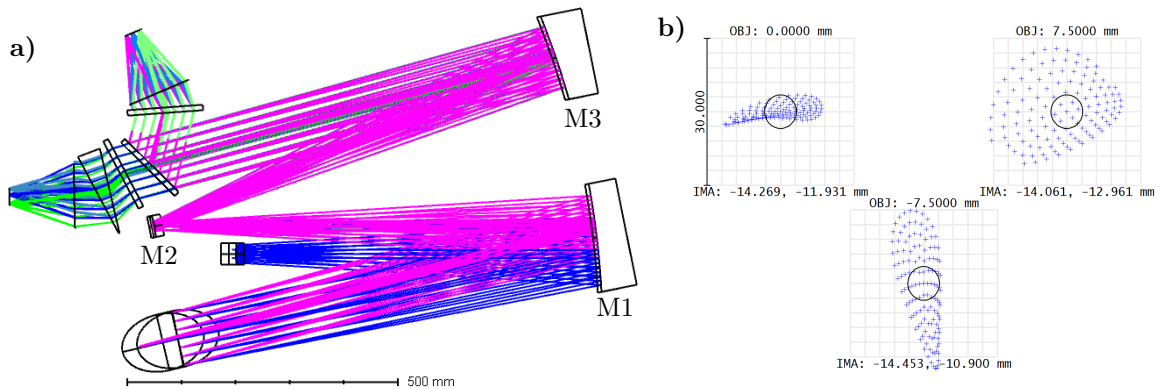
\begin{figure}[p]
    \centering
    \begin{overpic}[trim={140, 90, 140, 10}, clip, height=0.3\linewidth]{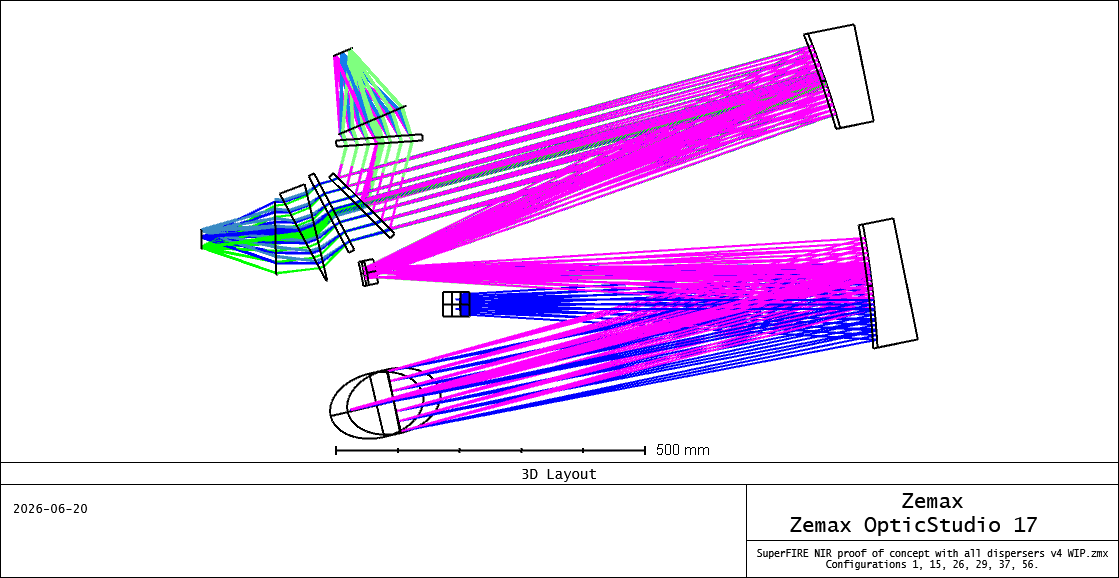}
        \put(2, 51){\bf a)}
        \put(92, 11){M1}
        \put(22, 19){M2}
        \put(87, 40){M3}
    \end{overpic}
    \begin{overpic}[trim={100, 250, 200, 10}, clip, height=0.3\linewidth]{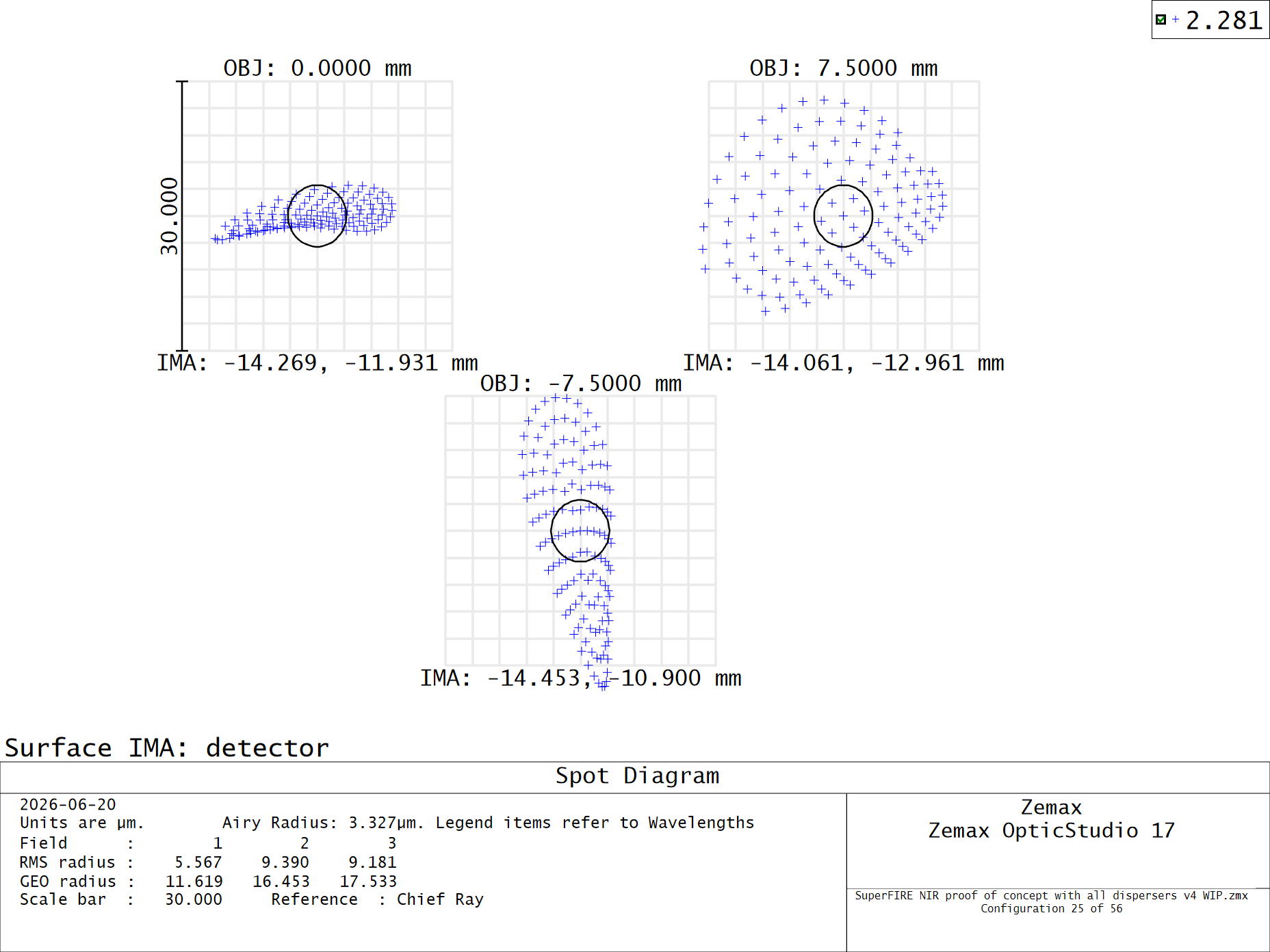}
        \put(0, 65){\bf b)}
    \end{overpic}
    \caption{Conceptual SuperFIRE NIR spectrograph layout (a) and image quality (b) at \SI{2.281}{\micro m} with paraxial cameras. A fold mirror feeds the light from a slit mechanism that mounts flat to the optical bench. A silicon prism is used in addition to a first-order grating for HK cross dispersion. The spot diagrams show a 2x2 pixel box.}
    \label{fig:spectrograph}
\end{figure}

The SuperFIRE baseline design uses Baranne's symmetrical white-pupil spectrograph\cite{baranneELODIESpectrographAccurate1996} with a Mangin mirror in place of a flat fold mirror to correct field curvature as introduced in G-CLEF\cite{fureszGCLEFSpectrographOptical2014}. In this configuration, shown in Figure~\ref{fig:spectrograph}a, an off-axis parabola (M1) collimates the light that passes through the slit and places a \SI{100}{mm} pupil on the echelle grating. The echelle is slightly tilted, so that the dispersed light is refocused by M1 onto the fold mirror (M2) adjacent to the slit and collimated by a second identical OAP (M3). The dispersed light overlaps in the ``white'' pupil that is formed after M3 so that the cross disperser and camera apertures are approximately the same size as the original beam diameter.

The UV-optical and NIR spectrographs use the same optical design and only need a different Mangin mirror optimized for the respective band. The image quality for this early design has a worst spot size of around 2 pixels EE80 diameter (Figure~\ref{fig:spectrograph}b) across all fields and wavelengths, which consumes the whole image quality budget. We are exploring evolutions of this design with a goal of reducing the spot sizes by a factor of 2 to leave most of the image quality budget for the cameras. As an example, we are considering adding correctors and/or using freeform mirrors in the current layout, or adding additional mirrors (e.g.\ replacing the M3 OAP with a three mirror anastigmat) to get the required wavefront control. The detailed optical design work on SuperFIRE has recently started and is a collaboration between the MIT Kavli Insitute and MIT Lincoln Laboratory.

Camera designs will be challenging for SuperFIRE, which requires fast cameras (f/1) with wide fields of view (\SI{20}{\degree} diagonal), but the previous concept study already showed that an f/1 \SI{24}{\degree} YJ camera design is feasible\cite{simcoeConceptSeeinglimitedNearIR2016}. Other relevant designs such as those for TESS (f/1.4, \SI{34}{\degree}, \SIrange{.6}{1}{\micro m}) \cite{chrispOpticalDesignCamera2015} and MOSFIRE (f/.93, \SI{12}{\degree}, \SIrange{1}{2.4}{\micro m})\cite{mcleanDesignDevelopmentMOSFIRE2010} provide useful starting points and suggest feasibility. Due to limited glass selections the u and K band cameras are expected to be the most challenging to design. The wide camera field of view also means that the cross dispersers need to be tilted significantly to avoid the Littrow recombination ghost\cite{burghRecombinationGhostsLittrow2007} which will reduce throughput. The most difficult cross disperser will be in the HK arm, however, this cross disperser can be used in Littrow because the central wavelength of order 15 (\SI{1.9}{\micro\m}) is blocked by the atmosphere and gives a natural place to ``dump'' the recombination ghost.

\subsection{Telescope integration}
\begin{figure}[p]
    \centering
    \begin{overpic}[width=0.7\linewidth]{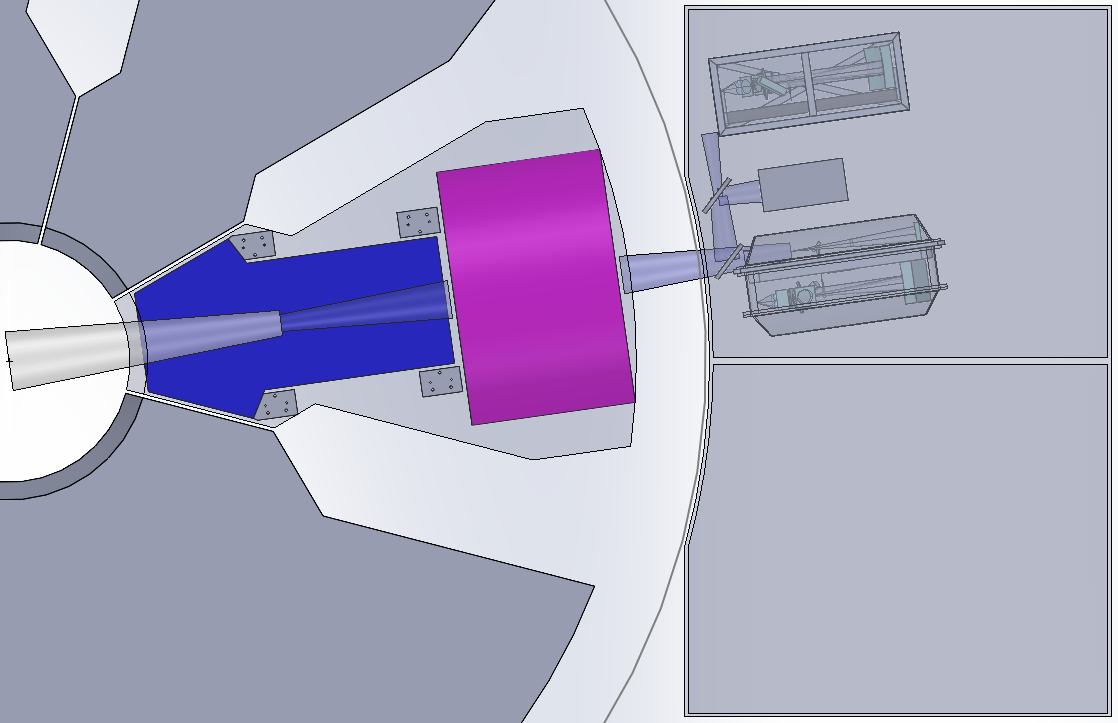}
        \put(40, 38) {\color{white} \bf \begin{tabular}{c} IP relay\\volume\end{tabular}}
        \put(15, 35) {\color{white} \bf \begin{tabular}{c} GMagAO-X\\ tertiary\end{tabular}}
        \put(78, 49) {\bf \begin{tabular}{c} SuperFIRE\\ location\end{tabular}}
    \end{overpic}
    \caption{Conceptual to-scale layout of SuperFIRE in the GMT IP station and the swept volume of the IP relay.}
    \label{fig:mockup}
\end{figure}

\begin{figure}[p]
    \centering
    \begin{overpic}[trim={100, 92, 100, 10}, clip, height=.33\linewidth]{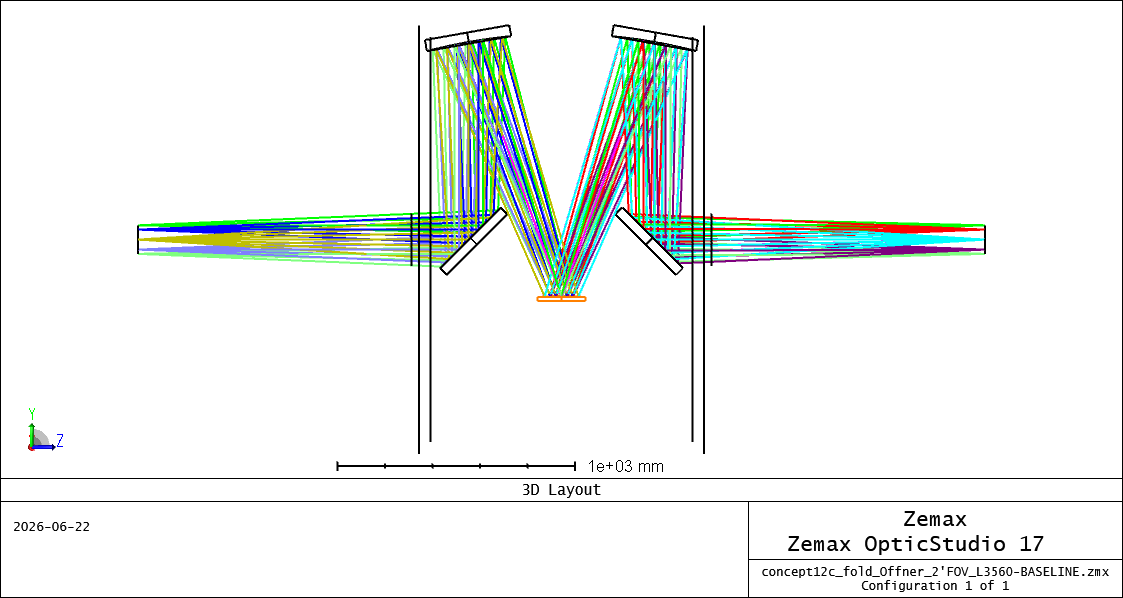}
        \put (1, 50) {\bf a)}
    \end{overpic}
    \begin{overpic}[trim={200, 100, 200, 150}, clip, height=.33\linewidth]{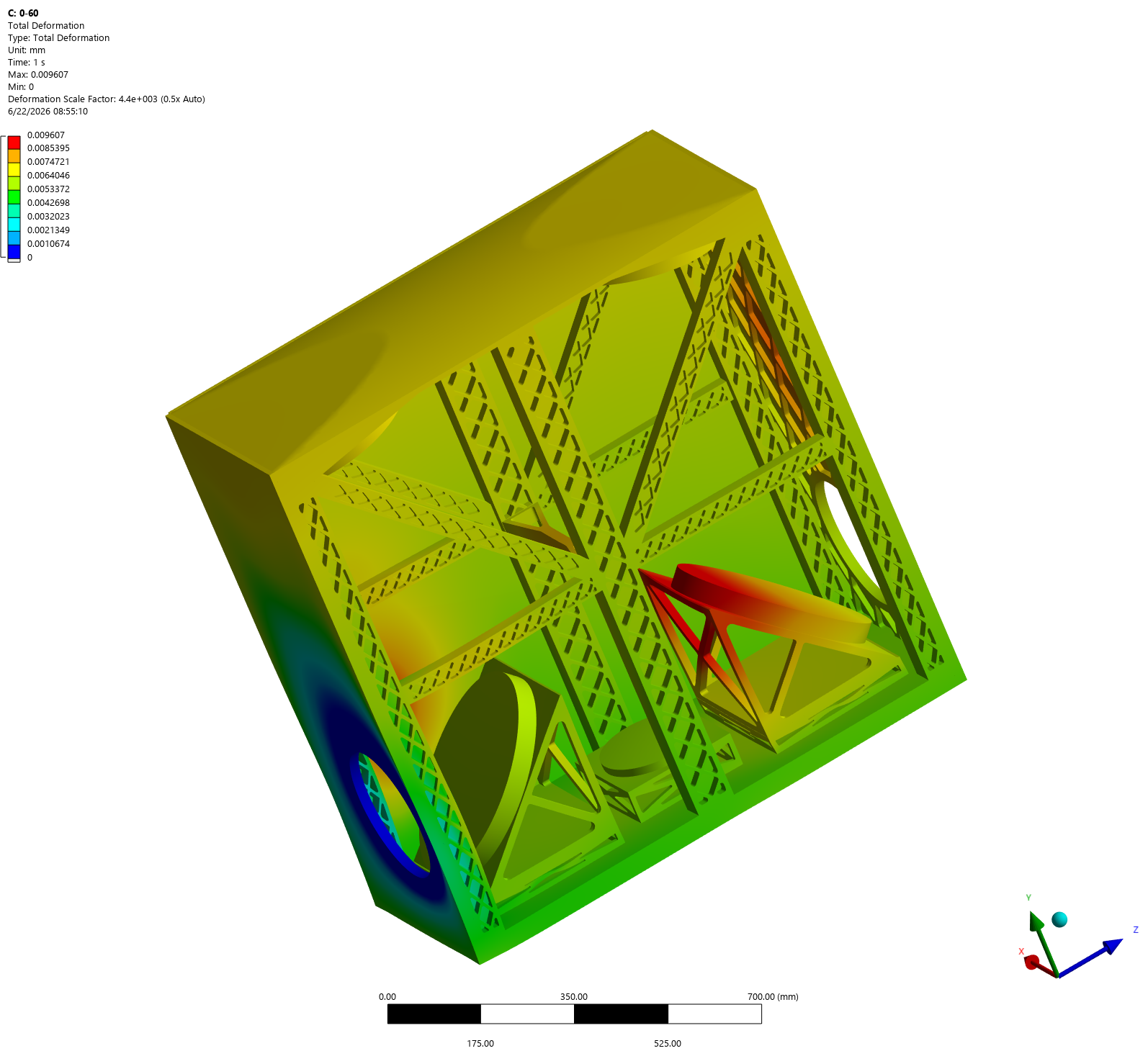}
        \put (1, 92) {\bf b)}
    \end{overpic}
    \caption{Conceptual design for the folded Offner-type IP relay. a) Ray-trace. b) Preliminary finite-element model of an all-aluminum structure with carbon fiber side panels showing a maximum deformation around \SI{10}{\micro m}.}
    \label{fig:relay}
\end{figure}

SuperFIRE is designed to occupy no more than half of the instrument platform (IP) station on the GMT where it will always be available for fast instrument switching. To use this station the GMT's Gregorian instrument rotator and facility tertiary are set to fixed positions and feed the IP relay system to provide a focus for the instrument, as shown in Figure~\ref{fig:mockup}. In our concept the IP relay is designed to relay and derotate the focal plane at least \SI{3.5}{m} away with a 2' FoV of which at least 1' is unvignetted. Due to the wide wavelength range we have baselined an all-reflecting design based on a ``folded'' Offner relay shown in Figure~\ref{fig:relay}. This design provides derotation with the minimal number of reflections by rotating the optical assembly, as we found that a non-derotating relay would need at least four mirrors plus a three-mirror derotator. With three spherical and two plane surfaces the design is relatively insensitive to alignment and affordable, and preliminary finite-element modeling suggests that reaching about \SI{10}{\micro m} deformations across the different gravity conditions is feasible. The image quality degradation is less than 10\% for the best-case .2" spot size and we are targeting a flat 90\% throughput by using broadband dielectric mirror coatings. 

\section{CONCLUSIONS}
We have evolved the SuperFIRE concept significantly from its original form in response to emerging scientific needs that we believe give long-term importance to seeing-limited broadband single-object spectrographs. With higher spectral resolution, faster low-noise detectors, and reduced backgrounds, this concept for SuperFIRE is well suited to follow up faint transients uncovered by Rubin and Roman and will perform better than the original concept for faint-object science cases. We have chosen to focus on a few technologies that can give dramatic performance enhancements and otherwise use traditional design concepts to balance risk and development time. We are also considering the practical operational aspects of the instrument in the concept development phase to ensure that overheads are kept to a minimum and that SuperFIRE will be a practical and versatile point-and-shoot instrument that provides consistent results.

\acknowledgments 
This work was enabled by The Kavli Foundation via grant PS-2025-GR-0237-3087. We would also like to thank Terry and Susan Ragon whose gift enabled MIT to join the GMT project as a founding institute and Curtis Marble for a gift supporting GMT instrumentation work at MIT.

\bibliography{report} 
\bibliographystyle{spiebib} 

\end{document}